\documentclass[twocolumn]{aastex631}
\usepackage{amsmath}

\newcommand{\jz}{\ensuremath{j_z}}
\newcommand{\jpar}{\ensuremath{j_{\parallel}}}

\newcommand{\bperp}{\ensuremath{b_{\perp}}}
\newcommand{\epar}{\ensuremath{e_{\parallel}}}

\newcommand{\jasym}{\ensuremath{j_{\rm a}}}

\newcommand{\awesom}{\texttt{aweSOM}}
\newcommand{\gsum}{$G_{\rm sum}$}

\newcommand{\ud}{\ensuremath{\mathrm{d}}}

\received{February 19, 2025}
\revised{April 18, 2025}
\accepted{May 2, 2025}

\submitjournal{ApJL}
\shortauthors{Ha et al.}

\begin{document}

\title{Machine-Learning Characterization of Intermittency in Relativistic Pair Plasma Turbulence: \\Single and Double Sheet Structures}

\correspondingauthor{Trung Ha}
\email{tvha@umass.edu}

\author[0000-0002-0786-7307]{Trung Ha}
\affiliation{Department of Astronomy, University of Massachusetts Amherst, Amherst, MA 01003, USA}
\affiliation{Center for Computational Astrophysics, Flatiron Institute, 162 Fifth Avenue, New York, NY 10010, USA}
\affiliation{Department of Physics, University of North Texas, Denton, TX 76203, USA}

\author[0000-0002-3226-4575]{Joonas Nättilä}
\affiliation{Department of Physics, University of Helsinki, P.O. Box 64, University of Helsinki, FI-00014, Finland}
\affiliation{Center for Computational Astrophysics, Flatiron Institute, 162 Fifth Avenue, New York, NY 10010, USA}
\affiliation{Physics Department and Columbia Astrophysics Laboratory, Columbia University, \\538 West 120th Street, New York, NY 10027, USA}

\author[0000-0002-2685-2434]{Jordy Davelaar}
\altaffiliation{NASA Hubble Fellowship Program, Einstein Fellow}
\affiliation{Department of Astrophysical Sciences, Peyton Hall, Princeton University, Princeton, NJ 08544, USA}
\affiliation{Center for Computational Astrophysics, Flatiron Institute, 162 Fifth Avenue, New York, NY 10010, USA}
\affiliation{Physics Department and Columbia Astrophysics Laboratory, Columbia University, \\538 West 120th Street, New York, NY 10027, USA}

\author[0000-0002-1227-2754]{Lorenzo Sironi}
\affiliation{Department of Astronomy, Columbia University, New York, NY 10027, USA}
\affiliation{Center for Computational Astrophysics, Flatiron Institute, 162 Fifth Avenue, New York, NY 10010, USA}

\begin{abstract}

The physics of turbulence in magnetized plasmas remains an unresolved problem. 
The most poorly understood aspect is intermittency---spatio-temporal fluctuations superimposed on the self-similar turbulent motions.
We employ a novel machine-learning analysis technique to segment turbulent flow structures into distinct clusters based on statistical similarities across multiple physical features. 
We apply this technique to kinetic simulations of decaying (freely evolving) and driven (forced) turbulence in a strongly magnetized pair-plasma environment, and
find that the previously identified intermittent fluctuations consist of two distinct clusters: 
\textit{i}) current sheets, thin slabs of electric current between merging flux ropes, and;
\textit{ii}) double sheets, pairs of oppositely polarized current slabs, possibly generated by two non-linearly interacting Alfv\'en-wave packets.
The distinction is crucial for the construction of realistic turbulence sub-grid models.

\end{abstract}

\section{Introduction} \label{sec:intro}

Turbulence is a seemingly chaotic mechanism occurring in fluids and plasmas \citep{davidson2004, biskamp_magnetohydrodynamic_2003}.
In practice, it enables a transfer of energy from a large scale $l_0$ onto smaller and smaller scales, $l_0 > l_1 > l_2 \ldots$, via the nonlinearities in the governing dynamical equations.
Classical theories of turbulence assume that this mechanism is volumetric and self-similar \citep{kolmogorov_local_1941}---that is, the shape of the turbulent eddies is identical on all scales $l < l_0$ within the inertial range.
On the other hand, experiments and numerical simulations have demonstrated that some structures in this flow become increasingly sparse in time and space as $l$ decreases \citep{she_universal_1994}; 
their distributions also become prominently non-Gaussian.
These spatio-temporal fluctuations of the turbulent flow can be broadly classified as \emph{intermittency}. 

We focus on plasma turbulence where both the bulk velocities, $\mathbf{u}$, and the internal magnetic field fluctuations, $\mathbf{b}$, co-interact \citep{iroshnikov_turbulence_1964, kraichnan_inertial-range_1965, goldreich_toward_1995}. 
Such plasma turbulence is ubiquitous in space physics and astrophysics, including, e.g., the solar corona \citep[e.g.,][]{parker_topological_1972, matthaeus_turbulence_1999},  interstellar medium \citep[e.g.,][]{larson_turbulence_1981, lithwick_compressible_2001}, and accretion flows around compact objects \citep[e.g.,][]{ripperda_magnetic_2020, nathanail_magnetic_2022}.
In these systems, intermittency plays a crucial role in controlling, e.g., (non-thermal) particle acceleration \citep[e.g.,][]{lemoine2023,nattila_radiative_2024}, cosmic ray transport \citep[e.g.,][]{fielding2023,kempski2023}, and magnetic field amplification \citep[e.g.,][]{galishnikova2022,sironi_generation_2023,beattie2024}.

The energy cascade by magnetized plasma turbulence is mainly controlled by the interaction between counter-propagating magnetic-shear-wave (Alfv\'en mode) packets \citep{goldreich_toward_1995}.
The nonlinear interaction between Alfv\'en waves is known to drive a secular energy transfer onto a higher-wavenumber Alfv\'en mode \citep{howes_alfven_2013,tenbarge2021}.
The generic picture has been confirmed in multiple numerical experiments \citep{nielson2013,ripperda2021,nattila2022}.
What has remained poorly understood is the nature of the intermittency---i.e., deviations from this picture of co-interacting Alfven wave packets.
One manifestation of such deviation could, e.g., be the speculated dynamical alignment between $\mathbf{u}$ and $\mathbf{b}$ fields \citep{boldyrev2006}, or, more accurately, between the related Els\"asser fields, $\mathbf{z}_\pm = \mathbf{u} \pm \mathbf{b}$ \citep{mallet2015, chernoglazov2021}.
Another possibility is to interpret current sheets (and their interactions) as a manifestation of the intermittency \citep{zhou_spectrum_2023}. 

We employ a theory-agnostic way of characterizing the intermittency in magnetically dominated plasma turbulence.
We study the nature of the intermittency in three-dimensional (3D) plasma turbulence using a combination of machine learning (ML) and computer vision techniques. 
Namely, we identify structures consistent with \textit{current sheets} via a non-linear $N$-dimensional correlation with a CPU/GPU-accelerated implementation of self organizing maps \citep[SOM;][]{kohonen_self-organizing_1990}. 
Then, we combine multiple realizations of the SOM model via a statistical ensemble learning framework to obtain statistically more reliable clustering solutions.
Intermittent structures in plasma turbulence have previously been identified and their statistics analyzed with the structure functions \citep{davis2024}, thresholding of current densities \citep{zhdankin_statistical_2013, kadowaki_mhd_2018}, Gaussian-fitting techniques \citep{greco_statistical_2009, chhiber_clustering_2020}, and, recently, ML-based segmentation \citep{serrano_scale_2025}.
Previous attempts at leveraging ML methods to detect current sheets include the $K$-means clustering and DBSCAN methods \citep{sisti_detecting_2021}, convolutional neural networks \citep{hu_identifying_2020}, and SOM \citep{bussov_maarja_segmentation_2021,kohne_unsupervised_2023, edmond_clustering_2024}.
These studies offer a promising outlook on the usability of novel computational methods to identify, in particular, current sheets in 2D simulations and in magnetospheric observations. 
Our study is the first of its kind that applies such technique to 3D simulations of highly-magnetized, relativistic plasma turbulence.

With this technique, we analyze the intermittent structures in the magnetically-dominated regime to find two distinct types of current sheets: single current sheets where magnetic structures of the same polarity are merging, and double current sheets where magnetic structures of opposite polarities are interacting.
Previous studies have mainly identified the single sheets \citep[e.g.,][]{zhdankin_statistical_2013,sisti_detecting_2021,davis2024} with some rare exceptions discussing the origin and formation of double-sheet structures \citep[e.g.,][]{howes_alfven_2013, howes_spatially_2018}.
Differentiation between the single and double sheets might have important consequences for understanding of the physics of dissipation in magnetized turbulence.

In Section~\ref{sec:simulation}, we describe the kinetic plasma simulation used in our analysis. We describe the ML method and its implementation in Section~\ref{sec:ml-method}. In Section~\ref{sec:result}, we report on the model's clustering results and on its implications in terms of the two distinct intermittent structures. In Section~\ref{sec:discussion}, we discuss our findings and implications for future research.

\section{Simulation} \label{sec:simulation}

\begin{figure*}[t!]
    \includegraphics[width=\linewidth]{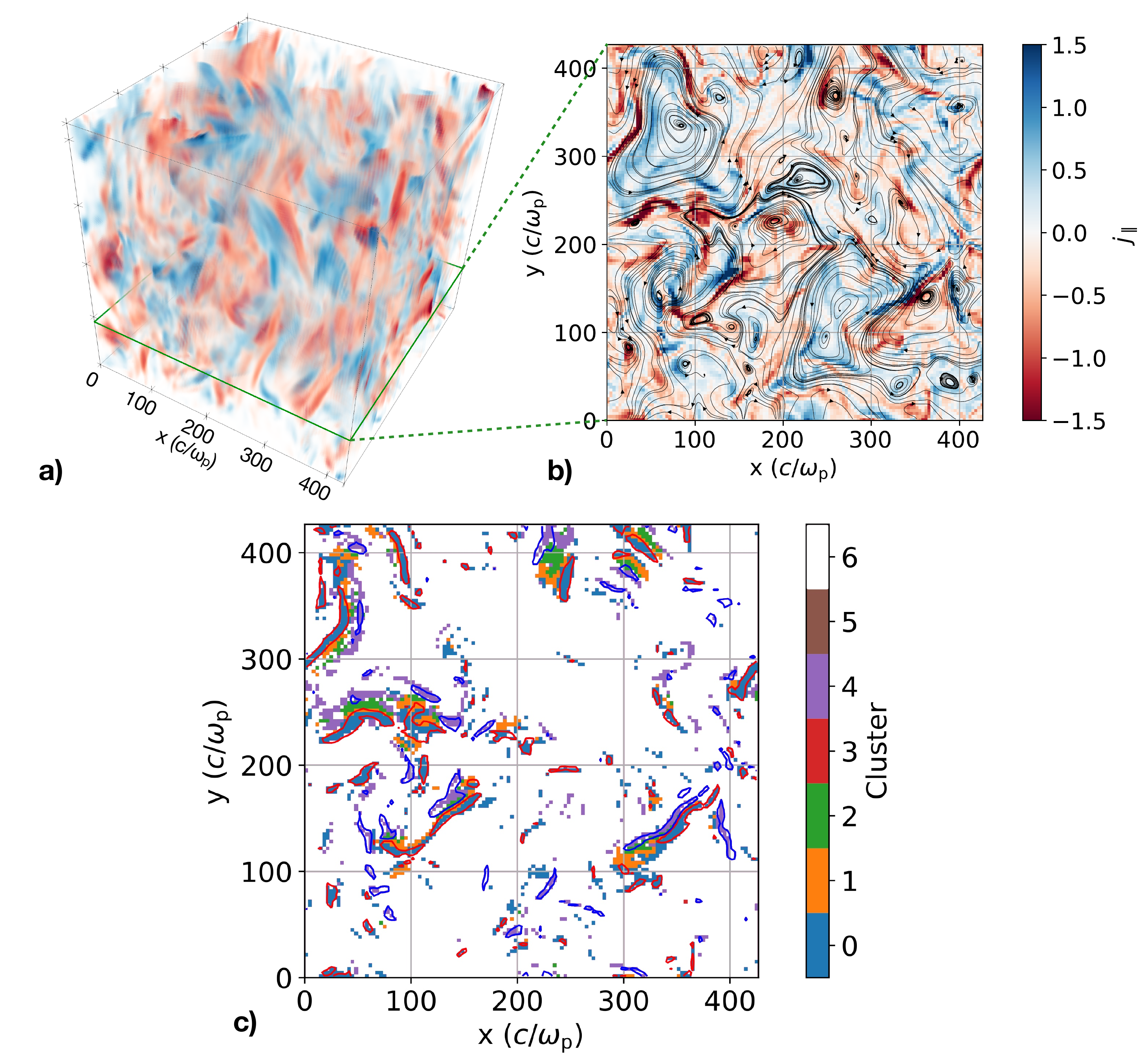}
    \caption{Panel a): volume rendering of the projection of the current density onto the $\mathbf{B}$~field, \jpar. Panel b): $xy$-slice of \jpar~at $z=90\, c/\omega_{\rm p}$. Streamlines of the in-plane magnetic field \bperp\ are in the foreground. Panel c): fiducial SOM segmentation, color-coded by the cluster identified via \awesom. From observation and the radar charts in Figure~\ref{fig:cluster_umap}, green and orange correspond to regions in between double current sheets, purple to current sheets aligned with $\mathbf{B}$, and blue to current sheets anti-aligned with $\mathbf{B}$. Blue contours are regions where $j_{\parallel} > 2 \, j_{\parallel, \rm rms}$, and red contours are regions where $j_{\parallel} < -2 \, j_{\parallel, \rm rms}$.
    }
    \label{fig:jz}
\end{figure*}

We analyze direct numerical simulations of 3D plasma turbulence.
We model the first-principle dynamics of the plasma with fully kinetic particle-in-cell (PIC) simulations \citep{birdsall} using the open-source \textsc{runko} code \citep{runko}.
Here, we report results from a domain characterized by a sudden excitation of turbulence, followed by the development of freely evolving (decaying) turbulence within the plasma.
Such a situation may be expected in the magnetic flaring events close to the accretion flows around black holes and/or neutron stars \citep[e.g.,][]{nattila_radiative_2024}.
Additionally, we have also verified that similar conclusions hold with continuously forced (driven) systems;
see Appendix~\ref{app:other-sims} for the analysis.

The technical details of the simulation are described in \citet{nattila_radiative_2021}.
Briefly, the simulation is a cubic box of size $L_{\rm sim} = 426\;c/\omega_{\rm p}$, covering 1280 grid cells per dimension.
The freely evolving system is initially in an unperturbed equilibrium state with neutral $e^\pm$-pair plasma, which is magnetized with a uniform external field $\mathbf{B_0} = B_0 \hat{\mathbf{z}}$. 
We focus on the strongly magnetized domain where the magnetization parameter $\sigma \equiv B_0^2/4\pi n m_e c^2 \gtrsim 1$ (where $n$ is the plasma number density, $m_e$ is the electron rest mass, and $c$ is the speed of light).
We set $\sigma=10$.
We consider the simplest pair plasma composition to minimize the role of the kinetic effects, with 54 particles per $c^3/\omega_{\rm p}^3$ per species (two particles per cell per species).
The initial equilibrium is disturbed by exciting large-scale ($=l_0 = L_{\rm sim}/4$) magnetic perturbations, $\delta \mathbf{B} \perp \mathbf{B_0}$ with a large amplitude $\delta B/B_0 \sim 1$. 
The resulting plasma bulk motions are trans-relativistic, since $|\mathbf{u}| \approx v_A \approx c$, where $v_A \equiv c \sqrt{\sigma/(\sigma+1)}$ is the Alfv\'en velocity.
We follow the simulations until $t = 20\, l_{0}/c$ (where $l_0/c$ is the eddy-turnover time of the energy-carrying scale). 
We store data snapshots (composed of $\mathbf{E}$, $\mathbf{B}$, and $\mathbf{J}$ fields) on a cadence of $\Delta t \approx 1 \, l_0/c$.
Numerically, each snapshot we analyze is a down-sampled rectangular cube composed of $L^3 = 128^{3} \approx 2 \times 10^6$ data points. 
Figure~\ref{fig:jz}a shows a volume rendering of the projected current density along the $\mathbf{B}$ field, $j_{\parallel} \equiv \mathbf{J}\cdot\mathbf{B}/(Bn_0ec)$, together with a $xy$-slice plot of the same feature, over-plotted by streamlines of the in-plane magnetic field, $B_{\perp}$ in Figure~\ref{fig:jz}b.
We present the following results based on one representative snapshot of the simulation, taken at $t = 5\;l_0/c$, which exhibits a magnetic power spectrum  $\propto k_\perp^\alpha$ with a slope of $\alpha \approx -5/3$, where $k_\perp$ is the wavenumber perpendicular to the mean magnetic field.

\section{Clustering Analysis} \label{sec:ml-method}
We analyze the resulting large, multidimensional dataset with a ML technique called Self-Organizing Maps \citep[SOM;][]{kohonen_self-organizing_1990}. 
SOM is a sophisticated classification method that can capture multidimensional correlations in the input data by exposing the topology of the (possibly nonlinear) manifolds being analyzed. 
We employ the SOM algorithm specifically because it is a powerful, unsupervised method that does not require prior knowledge about the number of clusters (i.e., \textit{physical structures}). Furthermore, the resulting clustering of nodes in a SOM is easily interpretable. 
We further enhance the robustness of the model by combining multiple SOM realizations through a statistically combined ensemble method \citep[SCE;][see also Appendix~\ref{app:sce}]{bussov_maarja_segmentation_2021}. SCE segmentation provides a statistically significant result by stacking SOM clusters of similar spatial distributions.
We provide a numerically fast implementation of the SOM and SCE methods as an open-source Python package called \awesom\footnote{\url{https://github.com/tvh0021/aweSOM/}} \citep{ha_awesom_2025}. 
Specifically, the SOM implementation in \awesom\ is an optimized and parallelized version of the R package \texttt{POPSOM} \citep{hamel_vsom_2019}, providing a marked improvement in training time, and vastly superior projection (the mapping of cluster label from lattice space to real space) time, on the order of $\sim 20-50$ times faster.

\begin{figure}[t]
    \centering
    \includegraphics[width=\columnwidth]{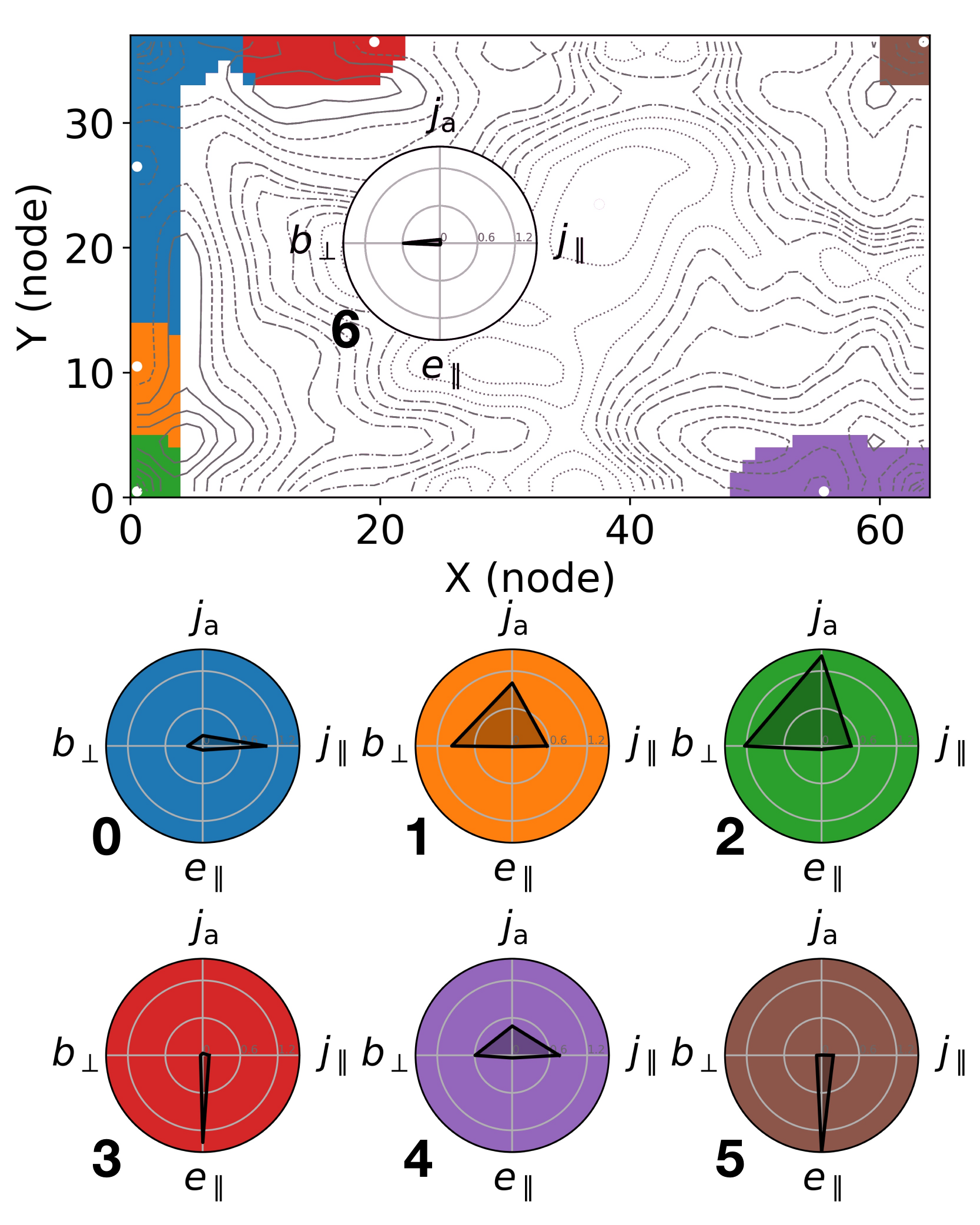}
    \caption{Fiducial SOM representation of the plasma turbulence data with unified distance matrix (U-matrix). 
    Each pixel in the map corresponds to a node in the lattice; the nodes are color-coded by their final cluster assignment. 
    Cluster centroids are marked with white circles. In this example, seven clusters are identified.
    The map is overlaid with U-matrix isocontours, visualizing the lattice-space distance between the nodes. The style of the isocontours represents the value of the U-matrix in ascending order: dotted lines, dotted-dashed lines, dashed lines, and solid lines, respectively.
    The radar charts visualize the relative importance of each feature of the given cluster.
    }\label{fig:cluster_umap}
\end{figure}

For our fiducial analysis, we utilize four features to train the model: 
\begin{enumerate}
    \item projection of the current along the magnetic field, $j_\parallel \equiv \mathbf{J}\cdot \mathbf{B}/(B n_0 e c)$,
    \item anti-symmetric convolution of the parallel current, $j_\mathrm{a} (\mathbf{x}) \equiv \int K_\mathrm{a}(\mathbf{y}) j_\parallel (\mathbf{x} - \mathbf{y}) \,\ud x \, \ud y$, where $K_\mathrm{a}$ is a two-dimensional anti-symmetric kernel (a discrete 1D counterpart is $\hat{K}_\mathrm{a}^{1\rm D} \equiv [-1, 0, 1]$),
    \item amplitude of the magnetic field in the $xy$-plane, $b_{\perp} \equiv b_{xy} \equiv (\mathbf{B} - \mathbf{B} \cdot \mathbf{\hat{z}})/B_0$,
    \item projection of the electric field along the magnetic field, $e_\parallel = \mathbf{E}\cdot\mathbf{B}/B_0$.
\end{enumerate}
Here, the anti-symmetric kernel $K_\mathrm{a}(r,\phi)$ (i.e., a matrix in our case) resembles the Heavyside-like function in a cylindrical coordinate system (as a function of $r$) and has, therefore, the property that 
$K_\mathrm{a}(r < r_\mathrm{cut}, \phi \in [-\pi/2,\pi/2]) = 1$, 
$K_\mathrm{a}(r < r_\mathrm{cut}, \phi \not\in [-\pi/2, \pi/2]) = -1$, 
$K_\mathrm{a}(r > r_\mathrm{cut}, \phi) = 0$, 
where $r_\mathrm{cut} = 3 c/\omega_{\rm p}$ is the kernel's extent, and $\phi$ is an angle against the direction set by $\nabla j_{\parallel}$.
In practice, if there is a location in the turbulent medium with doubly-peaked but oppositely directed current distributions with peaks of $-j_\mathrm{peak}$ and $+j_\mathrm{peak}$ next to each other in the $xy$-plane, then $j_{\rm a} \approx 2 j_\mathrm{peak}$ in between the peaks;
if the peaks are of same polarity, $+j_{\mathrm{peak},1}$ and $+j_{\mathrm{peak},2}$, then $j_{\rm a} \approx 0$.
Meanwhile, \jpar, \bperp, and \epar\ describe the magnitude of the current (which serves as a proxy for the current sheets), perpendicular magnetic field (a proxy for magnetic fluctuations), and non-linear electric field (a proxy for invalidity of ideal magnetohydrodynamics), respectively.

We normalize the input data before training such that each feature has a distribution with a mean $\mu = 0$\ and standard deviation $\sigma = 1.0$ (see Appendix~\ref{app:convergence} for a more detailed discussion).
With this approach, the four features are weighted equally by the model, while the outliers within each distribution are preserved, thus highlighting regions that experience strong intermittency.

Our fiducial SOM has a size of $64\times 37$ nodes in the lattice space. 
The map is trained with an initial learning rate $\alpha_0 = 0.1$ for $N = L^3 \approx 2\times10^6$ steps.
We have found the reported map size to be the minimum viable option for adequately resolving the multidimensional feature space.
We use a map with an aspect ratio of $H = 37/64 \approx 0.6$.
The aspect ratio is selected to be close to the ratio of the first and second largest eigenvalues of the input data's principal components. 
A detailed discussion of the technical parameters is given in Appendix~\ref{app:som}.
During our extensive tests (see Appendix~\ref{app:convergence} and \ref{app:distribution}, we found that the SOM lattice encompasses a surface in the 4D feature space, which is not easily translated to a physical connection. 
For example, Figure~\ref{fig:lattice_triangle} shows projections of the lattice on pairs of features.
The nodes along $Y=0$ (the black line going from green to purple) span a wide range of values in \jasym, but does not show a similar behavior in any other feature.
At the same time, the nodes along $Y=36$ (the black line going from blue to brown) span both extremes in \epar.
The shorter lattice direction ($Y$), on the other hand, tends to separate the positive and negative data values;
physically, this can be interpreted as separating the directions aligned and anti-aligned with $\mathbf{B_0}$.

\section{Results} \label{sec:result}

\subsection{Fiducial SOM realization}

Figure~\ref{fig:cluster_umap} shows the unified distance matrix (U-matrix) of the trained SOM as isocontours. 
The unified distance corresponds to the feature-space distance between adjacent nodes in the lattice. 
For our fiducial case, the SOM detects seven clusters of distinct multi-dimensional correlations between the provided features. 
We conclude that:
\begin{itemize}
\item Cluster 0 (blue in Figure~\ref{fig:cluster_umap}) shows a strong correlation with \jpar.
This is clearly seen when compared with a \jz~slice. 
The cluster highlights regions with strong current density but where the direction is anti-aligned against $\mathbf{B}$.
\item Clusters 1 (orange) and 2 (green) have relatively strong \bperp~and \jasym, indicating the regions between double current sheets.
Because the y-axis of the U-matrix indicates the alignment with the direction of $\mathbf{B}$, cluster 1 is anti-aligned, while cluster 2 is aligned with $\mathbf{B}$. 
\item Cluster 3 (red) and cluster 5 (brown) have exceptionally strong \epar.
These are regions where $\mathbf{E}$ strongly aligns with $\mathbf{B}$, indicating a breakdown of the (ideal) MHD approximation. Value-wise, cluster 3 indicates an anti-alignment with $\mathbf{B}$, while cluster 5 indicates an alignment with $\mathbf{B}$.
\item Cluster 4 (purple) is similar to cluster 0, but in these regions \jpar~align with $\textbf{B}$. There is a weak dependence on \bperp~and \jasym~that are more noticeable than in cluster 0, but this region is still representative of current sheets. 
\item Lastly, cluster 6 (white) coincides with non-active background plasma regions and is thus of little physical interest to us here.
\end{itemize}

Altogether, clusters 0 and 4 are single current sheets in isolation, but when detected adjacent to clusters 1 and 2, they are components of double current sheets. 
Taking a $xy$-slice as seen in Figure~\ref{fig:jz}b, we point to the structure at $x = 90\, c/\omega_{\rm p}$\ and $y = 400\, c/\omega_{\rm p}$\ as an example of a single current sheet, while at $x = 350\, c/\omega_{\rm p}$\ and $y = 150\, c/\omega_{\rm p}$\ a double current sheet is found.
The volume-filling fraction, $n_{\rm fill}$, of these clusters are (for cluster 0, 1, $\ldots$): $5.4\%$, $1.7\%$, $1.1\%$, $2.5\%$, $3.3\%$, $1.5\%$, $84.5\%$. 
The regions with sheet-like structure contains a volume-filling fraction of $n_{\mathrm{fill,\, single+double}} \approx 11.5\%$ of the domain, which is in line with similar studies of strongly magnetized plasma turbulence \citep[e.g.,][]{tenbarge_current_2013,vega_electron-scale_2023}.
An additional statistical breakdown of each cluster is examined in Appendix~\ref{app:distribution}.

\subsection{Separating single and double current sheets}
\label{subsec:separate}

\begin{figure}[t]
    \centering
    \includegraphics[width=\columnwidth]{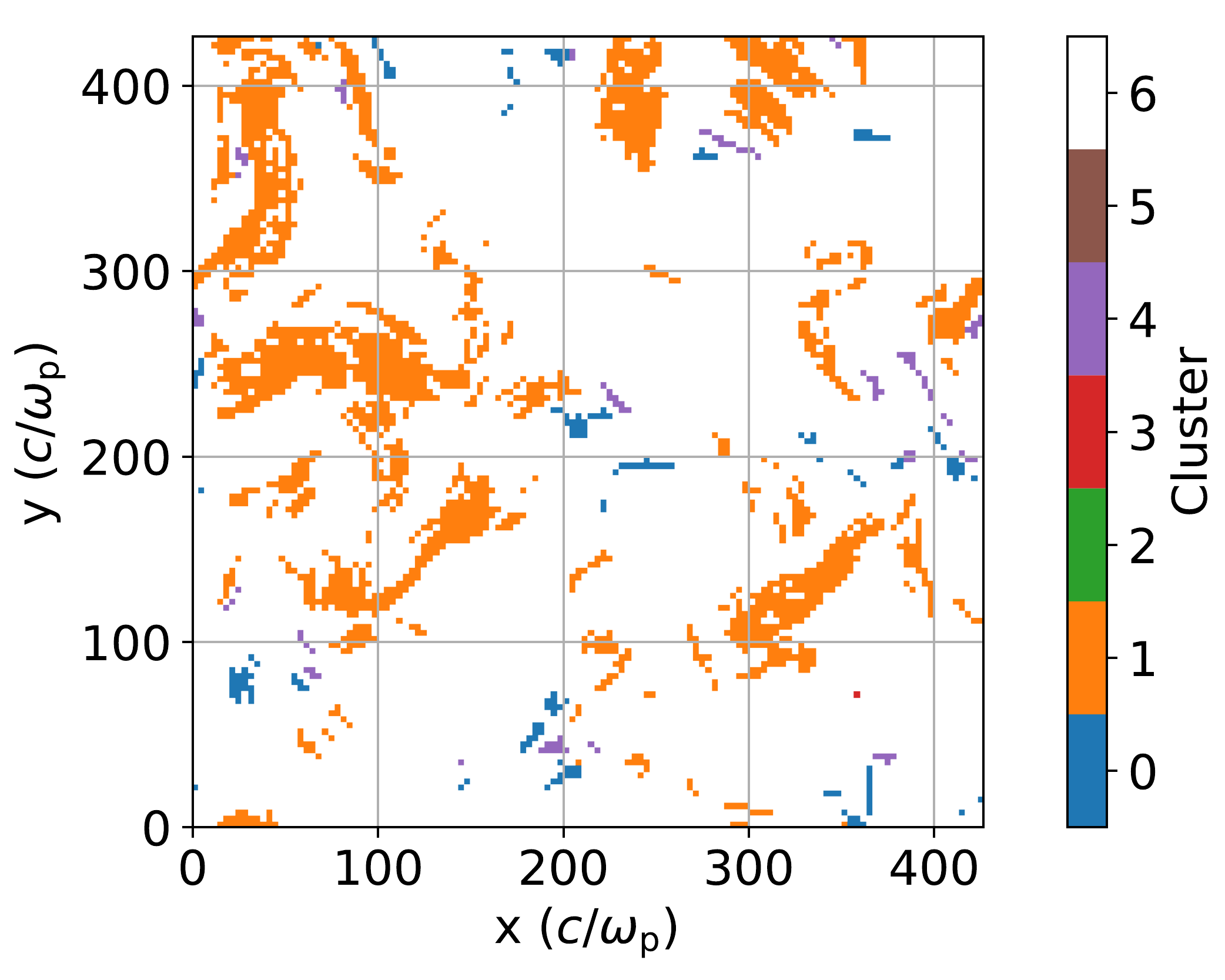}
    \caption{Similar to panel c) of Figure~\ref{fig:jz}, but with adjacent clusters 0, 1, 2, and 4 merged into cluster 1. Cluster 1 denotes double current sheets, cluster 4 and 0 denotes single current sheets that align with \textbf{B} and anti-align with \textbf{B}, respectively.}
    \label{fig:merged_double}
\end{figure}

From the fiducial snapshot shown in Figure~\ref{fig:jz}c, we compute the volume-filling fraction of the single and double current sheets separately. 
Since cluster 1 and 2 only occupy the spaces in between peaks of the double current sheets, we merge all continuous double sheets clumps with their adjacent single current sheets (cluster 0 and 4) into complete double current sheets. 
The remaining clumps from cluster 0 and 4 that are not merged are true single current sheets. 
Lastly, clumps identified as cluster 1 or 2 but are not found to be adjacent to both a $-j_{\rm peak}$ and a $+j_{\rm peak}$ are considered false detections, and are merged into the background cluster. 

Figure~\ref{fig:merged_double} shows an $xy$-slice of the clustering result after merging, keeping the same color notations as in Figures \ref{fig:jz}c and \ref{fig:cluster_umap}.
Here, cluster 1 (orange) represents double current sheets; cluster 4 (purple) and cluster 0 (blue) are single current sheets that are aligned and anti-aligned with \textbf{B}, respectively; cluster 6 (white) is the background plasma; cluster 3 (red) and 5 (brown) are still locations of strong alignment between \textbf{E} and \textbf{B}.
Altogether, $n_{\rm fill,\, double} \approx 9.1\%$\ and $n_{\rm fill,\, single} \approx 1.5\%$. 
When the system is allowed to reach steady state without continuous energy injection,
we find that most ``current sheet'' structures found are double current sheets. On the other hand, in continuously-driven turbulent environments, the relative fraction of single current sheets is more comparable to that of double current sheets ($n_{\rm fill,\, double} \approx 0.91\%$\ and $n_{\rm fill,\, single} \approx 0.46\%$ in our analysis; see Appendix~\ref{app:other-sims}).

\subsection{SCE result}

\begin{figure*}[t!]
    \centering
    \includegraphics[width=\linewidth]{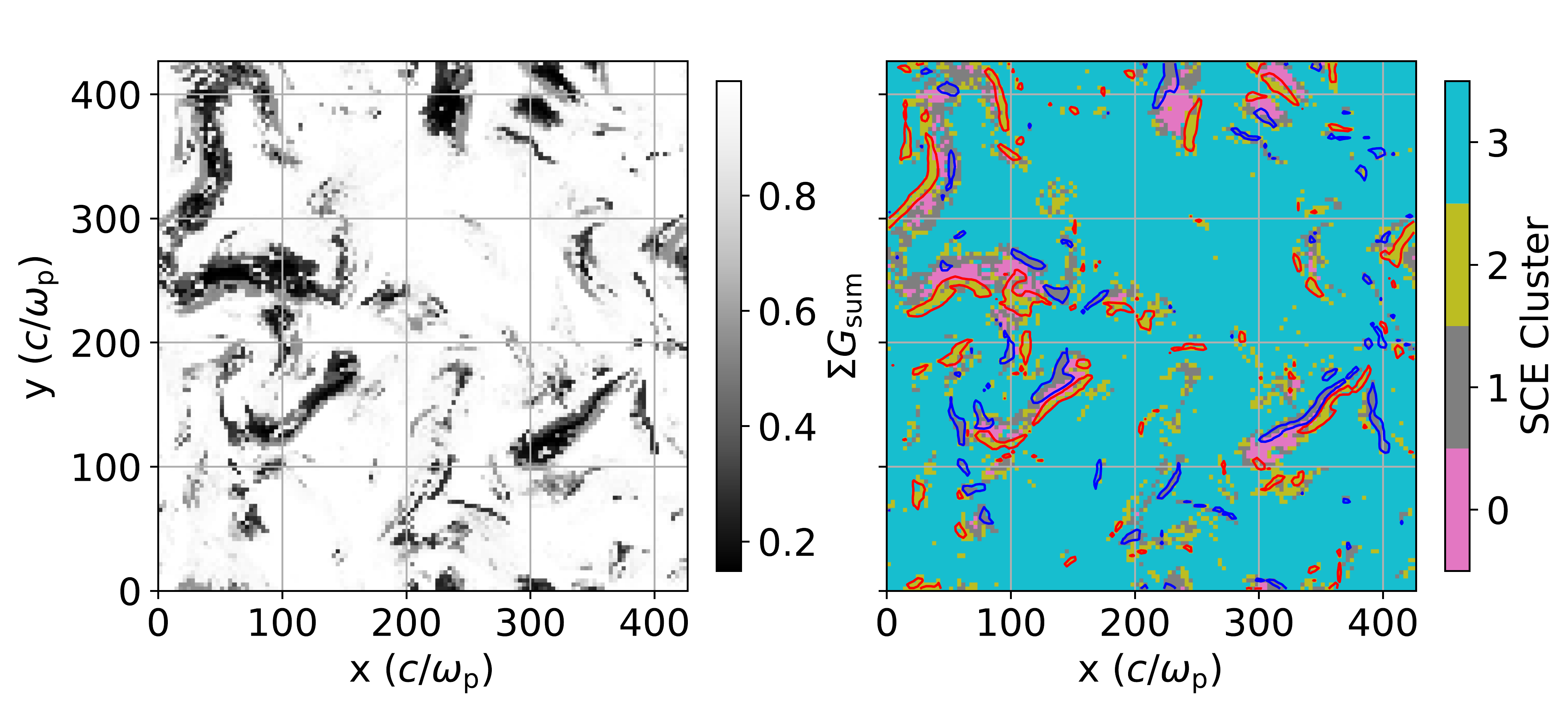}
    \caption{SCE result obtained by combining the output of 36 SOM realizations. 
    The left panel shows a $xy$-slice plot of $\Sigma G_{\rm sum}$ (signal strength) at $z = 90\, c/\omega_{\rm p}$. 
    The right panels show SCE clusters, obtained via making thresholds in the signal strength at 0.25, 0.5, and 0.8, respectively. Red and blue contours are the same as in Figure~\ref{fig:jz}c, and denotes the regions where $|j_{\parallel}| > 2 \, j_{\parallel, \rm rms}$.
    Nominally, cluster 0 (pink) are double current sheets, cluster 1 (gray) are single current sheets aligned with $\mathbf{B}$, cluster 2 (olive) are single current sheets anti-aligned with $\mathbf{B}$, and cluster 3 (teal) are background plasma.}
    \label{fig:SCE_clusters}
\end{figure*}

A key disadvantage of unsupervised ML methods compared to their supervised counterparts is the tendency of the final result to be highly dependent on the initial conditions \citep[e.g.,][]{attik_self-organizing_2005}. 
We find very similar sets of final clusters when slightly varying the SOM hyperparameters ($N$, $H$, and $\alpha_0$). The results are also robust against the selected input features or turbulence parameters. However, as more features are added to the model, one would naturally expect to find more intrinsic clusters in the domain. Nevertheless, the clusters that specifically point to intermittency are always present.
These observations strongly suggest a universality of the obtained clusters by nature of a high-dimensional correlation between the most important features (\jpar, \jasym, \bperp, and \epar).

We quantitatively verify the robustness of these clustering results by conducting an SCE analysis on 36 different SOM realizations. The SCE clusters are derived by combining the strongest common features across the SOM realizations—features that are more universally present are indicated by stronger signals. Figure~\ref{fig:SCE_clusters} shows a $xy$-slice plot of the cumulative SCE \gsum\ values obtained via cluster-to-cluster comparison of all clusters $C$ in all \texttt{aweSOM} realizations $R$ (see Appendix~\ref{app:sce} for a more detailed definition of \gsum). The prominent clusters reveal four main structures: individual current sheets in olive and gray, double current sheets in pink, and the background plasma in teal.
We caution that, since the SCE analysis is run on the original SOM clustering results without post-processing, many of the individual sheets identified by the SCE algorithm are not of single type, but instead parts of the double-sheet structures if detected adjacent to cluster 0 (pink).
The more accurate statistical representation of the two types of intermittent structures is presented in Section~\ref{subsec:separate}.

\begin{figure*}[ht]
    \centering
    \includegraphics[width=\linewidth]{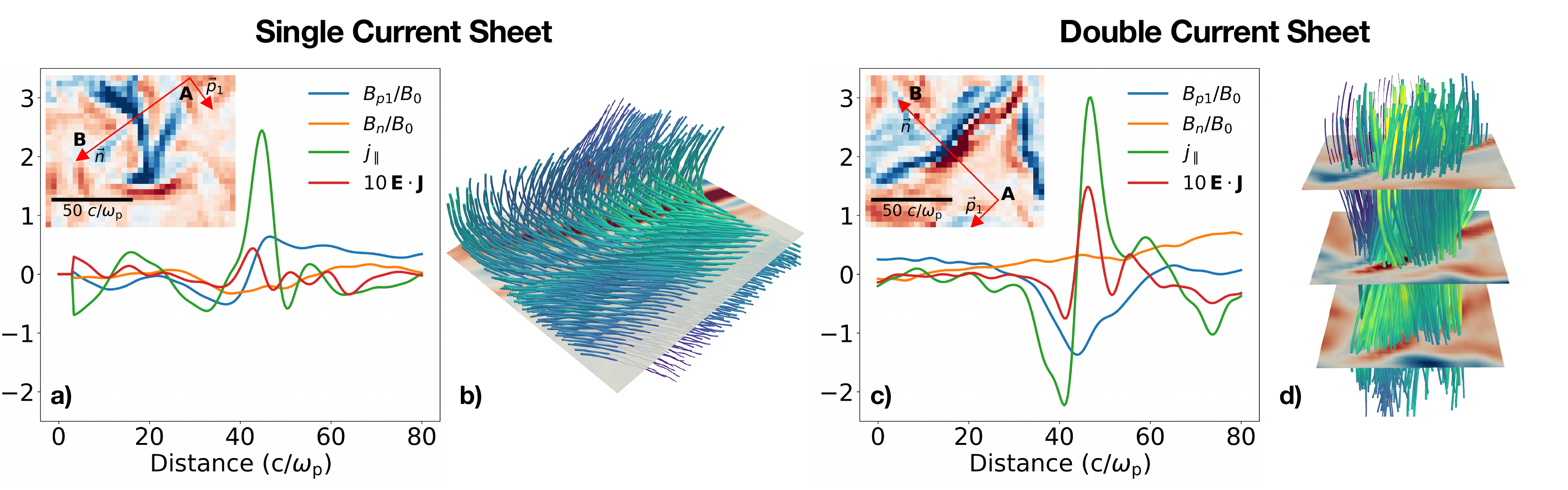}
    \caption{Panel a) and c): line profiles of \jpar, $B_{n}/B_0$, $B_{p1}/B_0$, and $\mathbf{E}\cdot\mathbf{J}$ along the line segment $\overline{AB}$. Each inset plot is a $xy$-slice of \jpar\ centered on the sheet of interest. 
    Panel b) and d): volume rendering of $\mathbf{B}$~field lines, interlaced with $xy$-slices of \jz. On the left is an example of a single current sheet; on the right is an example of a double current sheet.
    }
    \label{fig:single_double}
\end{figure*}

\subsection{Intermittency and current sheets}%
Interestingly, we find that some of the identified clusters are associated with spatio-temporal fluctuations in the simulation domain (clusters 0, 1, 2, and 4 in our fiducial analysis visualized in Figure~\ref{fig:jz}c and \ref{fig:cluster_umap} and cluster 0, 1, and 2 in Figure~\ref{fig:SCE_clusters}).
Physically, these clusters coincide with localized patches of intense electric current: \emph{current sheets}.
Because of this, we associate these clusters with MHD intermittency.
However, we always find two differing sheet-like structures in the flow.
This is in stark contrast with the previous analyses
\citep[e.g.,][]{chernoglazov2021,fielding2023,davis2024}
that have assumed all current sheets to be of the same origin.

Figure~\ref{fig:single_double} shows an example of the two types of current sheets we regularly identify with the SOM analysis. 
The inset plot in the top panels shows a $xy$-slice plot of \jpar, overplotted with a line segment $\overline{AB}$, which is approximately normal to the sheet(s). 
We denote the direction along $\overline{AB}$ as $\mathbf{n}$, and the direction along $\mathbf{\hat{z}} \times \mathbf{n}$ as $\mathbf{p_1}$. 
Then, we compute the line profile of \jpar, $B_{p_1}$, $B_n$, and $\mathbf{J} \cdot \mathbf{E}$~along $\overline{AB}$, and plot the profiles in panels a) and c).

The first type of cluster resembling intermittency is found in locations with large $j_\parallel$, small $j_\mathrm{a}$, and comparatively small $b_{xy}$ values (see clusters 0 and 4 in  Figure~\ref{fig:cluster_umap} and cluster 1 and 2 in Figure~\ref{fig:SCE_clusters})
We call these structures \emph{single sheets}. 
These single current sheets are classic exhibitions of intermittency via a reversal of magnetic polarity across the spatial domain.
Indeed, in Figure~\ref{fig:single_double}a, we find a relatively sharp peak in \jpar\ across a thickness of $\sim$$5 c/\omega_{\rm p}$. 
A clear reversal in the magnetic polarity is also observed at this peak, where $B_{p_1}/B_0$ goes from $-0.5$ to $0.5$, and there is no definitive change in $B_n$.
In a realistic, turbulent flow, these structures are composed of magnetic field lines rapidly changing their orientation by about $90$ degrees, i.e., a classical magnetic reconnection setup in a moderate-to-strong guide field regime (see  Figure~\ref{fig:single_double}b).

The second type of cluster resembling intermittency is found in locations with large $j_\parallel$, large $j_\mathrm{a}$, as well as large $b_{xy}$ values (see cluster 1 and 2 in Figure~\ref{fig:cluster_umap}, and cluster 0 in Figure~\ref{fig:SCE_clusters}). 
We call these structures \emph{double sheets} because they always come in pairs with oppositely directed currents. 
Figure~\ref{fig:single_double}c shows the line profiles across one double sheet. 
Importantly, rather than demonstrating a reversal in the magnetic polarity, there is a prominent peak in $B_{p_1}$ at the transition point between the sheets, indicating a sharp increase of the in-plane magnetic flux in the direction perpendicular to the sheet's surface normal.
In a realistic flow, these structures manifest as regions of the magnetic field with a separate, compressed core bundle oriented to a different direction than the neighboring field lines (see  Figure~\ref{fig:single_double}d).

\section{Discussion} \label{sec:discussion}
The formation of the single sheets is governed by the dynamics of the coherent structures in the turbulent flow \citep[e.g.,][]{zhou_spectrum_2023}.
First, we note that MHD turbulence tends to generate and sustain magnetic flux ropes (helical magnetic field structures with an electric current flowing along $\mathbf{B}$).
The dynamics between such structures follow from basic electrodynamics:
flux ropes with the same polarity attract; 
opposite polarities repel each other.
Mergers of the same-polarity flux ropes form a reversal in the $\delta\mathbf{B} \perp \mathbf{B}_0$ magnetic field components;
such region naturally induces a spatio-temporal current sheet.
Such single sheets are prone to magnetic reconnection 
\citep[e.g.,][]{priest_magnetic_2000,zweibel_magnetic_2009} and their lifetime is regulated by the tearing-instability \citep[e.g.,][]{comisso2019} (or their environment).
They are well-known sites of plasma energization \citep{sironi2014, comisso_particle_2018}.

Importantly, our analysis also demonstrates that plasma turbulence tends to host a second type of current sheets, which has been largely overlooked so far.
These double sheets have been seen in previous numerical studies \citep[e.g.,][]{servidio_magnetic_2009, zhou_spectrum_2023} and observations of the Earth's magnetosphere \citep[e.g.,][]{ergun_observations_2009}; however, their statistics has not been analyzed nor has their importance been highlighted in the past.
The physical origin of the double current sheets remains uncertain.
Most likely, the double sheets arise from the nonlinear interactions between Alfv\'en wave packets \citep{howes_alfven_2013,howes_dynamical_2016,verniero_nonlinear_2018}. 
In a freely-evolving relativistic pair plasma turbulence, we observe that the double current sheets are formed where magnetic structures of opposite polarities (i.e., rotational directions) are interacting.
However, even if their origin is not yet understood, their existence is important for future analysis of turbulence.
For example, it is not yet understood what scatters the energetic cosmic rays in the inter- and intra-galactic plasmas:
one hypothesis is the intermittent structures in the turbulence \citep{kempski2023, lemoine2023}.
A drastic difference in the shape of the (magnetic) structures can then lead to a significant difference in the scattering properties.
Lastly, the two intermittent structures also have differing dissipation profiles, which need to be explicitly accounted for in future turbulence sub-grid models that make assumptions about the local heating and particle acceleration.
For example, \citet{howes_spatially_2018} and \citet{nattila_heating_2022} found that Landau damping plays an important role in the energy dissipation of kinetic-Alfv\'en-wave turbulence.
Future studies on the stability of current sheets should also investigate the dissipation profile of these sheets, and whether they are analogous to the ``double tearing mode" current sheets in the literature \citep[e.g.,][]{pritchett_linear_1980,baty_explosive_2013}.

\begin{acknowledgments}
We would like to thank the anonymous reviewer for their constructive feedback which improves the clarity of this paper.
The authors thank Shirley Ho and Kaze Wong for useful discussions.
TH acknowledges support from a pre-doctoral program at the Flatiron Institute. 
Research at the Flatiron Institute is supported by the Simons Foundation.
JN is supported by an ERC grant (ILLUMINATOR, 101114623). 
JD is supported by NASA through the NASA Hubble Fellowship grant HST-HF2-51552.001-A, awarded by the Space Telescope Science Institute, which is operated by the Association of Universities for Research in Astronomy, Incorporated, under NASA contract NAS5-26555.
LS acknowledges support from DoE Early Career Award DE-SC0023015. This work was supported by a grant from the Simons Foundation (MP-SCMPS-00001470) to LS, and facilitated by Multimessenger Plasma Physics Center (MPPC), NSF grant PHY-2206609 to LS. LS is also supported by NSF grant PHY-2409223.
Simons Foundation is acknowledged for computational support.
\end{acknowledgments}

\software{JAX \citep{jax2018github}, numba \citep{lam2015numba}, matplotlib \citep{Hunter:2007}, numpy \citep{harris2020array}, \awesom \citep{ha_awesom_2025}}

\appendix

\section{SOM segmentation method} \label{app:som}
SOM is an unsupervised ML technique that uses competitive learning for tasks like dimensionality reduction, clustering, and classification \citep{kohonen_self-organizing_1990}. 
Fundamentally, a SOM is a 2D lattice of nodes that, through training, adapts to the intrinsic orientation of high-dimensional input data.

We develop a custom-made SOM implementation, \awesom\ \citep{ha_awesom_2025}. 
This Python package employs the same technique as similar Python-based SOM implementations. Specifically, we base \awesom\ on \texttt{POPSOM}\footnote{\url{https://github.com/njali2001/popsom}}.
We introduce optimizations to improve analysis speed, enabling it to handle complex, high-dimensional physical data. 
Below, we briefly describe \awesom's approach to learning the magnetized plasma turbulence.

First, we initialize a lattice of size $X\times Y \times F$, where $X$ and $Y$ are the number of nodes along each map direction, and $F$ denotes the number of features supplied to the model.
Before training, the initial weight value of each node, $\omega_0$, is randomly assigned, typically following a distribution representative of the input data.
The larger the lattice, the more details from the intrinsic data we can learn.
Kohonen \citep{kohonen_self-organizing_1990} advised using a lattice size of $N_{\rm node} = X \cdot Y = 5 \sqrt{N}$, where $N$ is the number of data points. We set $N_{\rm node} = \frac{5}{6} \sqrt{N \cdot F}$ such that the map both scales with the number of features in addition to the size of the data, but with a fraction of $\frac{1}{6}$ compensating for the map size quickly becoming too big to train when $N$ and $F$ are both large.

During each epoch, $t$, one input vector (a cell within the simulation domain) is randomly drawn. 
Then, the Euclidean distances, $D_{\rm E}$, between this vector and all nodes in the lattice are calculated.
The node with the smallest distance is chosen as the best-matching unit (BMU). 
Then, the weight value of each node is updated:

\begin{equation}
    w_{i,j}(t) = w_{i,j}(t-1) - D_{\mathrm{E}|i,j} \cdot \gamma(t),
\end{equation}
where $i,j$ represent the node's location in the lattice, and $\gamma(t)$ is the neighborhood function:
\begin{equation}
    \gamma(t)= \begin{cases}
        \alpha(t) e^{\frac{-d_{\rm C}^2}{2(s(t)/3)^2}}, & \text{if $d_{\rm C} \leq s(t)$}\\
        0, & \text{if $d_{\rm C} > s(t)$}
    \end{cases}
\end{equation}
where $\alpha(t)$ is the learning rate at epoch $t$, $d_{\rm C}$ is the Chebyshev distance between the BMU and the node at $(i,j)$, and $s(t)$ is the neighborhood width at epoch $t$.

At the core of the SOM technique is the shrinking neighborhood. 
Initially, $s_0 = \mathrm{max}(X,Y)$ such that earlier training steps adjust the weight values across the entire lattice.
As training progresses, $s$ gradually decreases until only a small number of nodes (or just the BMU) are updated each epoch. 
In \awesom, the final neighborhood size is set to $s_{\rm f} = 8$. 
This ensures that learning localizes to a specific region of the lattice without being overly restrictive, thereby preserving generalization.

After training, clustering is performed on the lattice based on the geometry of the U-matrix. 
Cluster centroids are identified by finding local minima in the U-matrix. 
A ``merging cost" is then calculated by line integration between all pairs of centroids.
If the cost is below a threshold of 0.25 (with the cost between the two furthest centroids on the lattice normalized to one), the clusters are merged. 
An example of the final clustering result on the lattice is shown in Figure~\ref{fig:cluster_umap}.

Lastly, the cells in the simulation are mapped to the nearest node in the lattice, each of which has been assigned a cluster label. 
This label is then transferred to the corresponding input vector, resulting in visualization of the clustering in the input space. 
An example of this mapping is shown in panel c) of Figure~\ref{fig:jz}.

\section{SOM convergence}\label{app:convergence}
We extensively test the convergence of the SOM clustering results by varying the key parameters before training: the number of training steps, $N$, the aspect ratio of the map, $H$, and the initial learning rate, $\alpha_0$.
Additionally, the choice of how to normalize the data before loading into \awesom, and how initial node values are assigned, can also contribute to the accuracy of the final clustering results.

We test the number of training steps required to reach convergence with $N$ from $10^4$~to $L^3$.
We find $N = 2\times10^5$~($\approx 10\% \; L^3$) to be on the lower bound for convergence, although higher $\alpha_0$~values (around 0.5) can allow convergence with fewer training steps.
Nevertheless, thanks to the relatively inexpensive cost of training a map, we choose $N=L^3=128^3$~for all cases.
Additional training by repeating the data does not appear to improve the results.

The initial learning rate $\alpha_0$ has a minimal impact on convergence, except when $N$ is small. During training, the learning rate is reduced at a constant rate such that the final learning rate $\alpha_{\rm f} = 10^{-3} \alpha_0$.
We use $\alpha_0 = 0.1$ for the fiducial realization but note that the U-matrices are qualitatively identical for $0.05 < \alpha_0 < 0.5$.

We also test the map's aspect ratio, $H$, by varying the ratio of vertical to horizontal nodes while keeping the total number constant.
Results are robust for $0.5 \lesssim H \leq 1$, and we use $H \sim 0.6$ in the fiducial realization, assuming a slight dominant of the data variance along one preferred direction.

We pre-process the data before training by normalization, testing three methods: \texttt{MinMaxScaler, StandardScaler} \citep{pedregosa_scikit-learn_2011}, and a custom normalization similar to \texttt{StandardScaler} but with a flexible range. 
In the end, we found that the custom normalization with a mean $\mu = 0$~and standard deviation $\sigma = 1.0$~for each feature is optimal for map generation.

We also test different methods for initializing the map. 
We choose the most general approach of drawing initial values from a uniform distribution with weights $-1 \leq w_0 \leq 1$. 
Alternatively, the lattice can be initialized by drawing random observations (cells in the simulation domain) and assigning those values as initial weights \citep{ponmalai_self-organizing_2019}. 
Ultimately, at $N = 128^3$, all initialization methods converge to similar final SOMs.

\section{SCE method} \label{app:sce}
\citet{bussov_maarja_segmentation_2021} discovered that their single SOM realizations were highly stochastic to small changes in the input parameters, raising concerns about the algorithm's stability.
To improve the robustness of these results, they developed an SCE method to stack multiple SOM realizations to obtain a single, markedly improved clustering result. 
The mathematical details of the SCE framework are discussed in \citet{bussov_maarja_segmentation_2021}. 
Below, we summarize the key concepts of SCE and how we integrate it into \awesom.

SCE involves a series of steps that stacks $n$~number of SOM realizations. For each cluster $C$ in a SOM realization $R$, its spatial distribution is compared with all other clusters $C'$ in $R' \neq R$ to obtain a goodness-of-fit index $g$. Then, each cluster $C$ is associated with a sum of goodness-of-fit (i.e. ``quality index"):
\begin{equation}
    G_{\rm sum} = \sum_{C_i' \in R'} g_i.
\end{equation}
Once all \gsum\ values are obtained, they are ranked in descending order, and groups of similar \gsum\ values are combined to form SCE clusters. 
This approach works because clusters with similar spatial distributions tend to have similar \gsum\ values \citep[see Figure 6 of][]{bussov_maarja_segmentation_2021}. In practice, we do not rank the \gsum\ values, but instead sum this index point-by-point to obtain a general ``signal strength" of each cell in the simulation. 

Single \awesom\ realizations of the 3D simulation are found to be robust when introducing small changes in the initial conditions (see the previous subsection). 
We independently confirm this robustness using SCE. 
Since this stacking process involves extensive tensor multiplications, we utilize GPU-accelerated computation with \texttt{JAX} \citep{jax2018github} and integrate this capability with \awesom. 
As a result, \awesom\ can perform SCE analysis on high-resolution ($N \gtrsim 1000^3$) 3D simulations using one GPU (we ran the SCE analysis on an NVIDIA V-100 with 32 GB of VRAM). 
The final SCE clustering result is shown in Figure~\ref{fig:SCE_clusters} and discussed in more details the next section.

\section{SCE clustering result} \label{app:sce-result}
We perform SCE analysis on a set of 36 SOM realizations, exploring a parameter space of $10^6 \leq N \leq 4 \times 10^6$, $0.4 \leq H \leq 1$, and $0.1 \leq \alpha_0 \leq 0.4$. 
We identify four statistically significant clusters. Figure~\ref{fig:SCE_clusters} highlights the prominence of each cluster as a cumulative sum of the number of pairs across the different SOM realizations that are detected at the same location. 
In the left panel, each pixel is color-coded based on its normalized $\Sigma$\gsum\ (signal strength). 
It is immediately clear that the background plasma has the highest signal strength ($\Sigma G_{\rm sum} > 0.8$) due to its high volume-filling fraction. 
The intermittent structures are found at lower signal strengths, with the single current sheets at the intermediate level ($0.25 \lesssim \Sigma G_{\rm sum} \lesssim 0.8$), and the double current sheets at the lowest signal strengths ($\Sigma G_{\rm sum} \lesssim 0.25$).
After setting three thresholds at $\Sigma$\gsum\ at 0.25, 0.5, and 0.8, we can point to cluster 0 as the double current sheets, cluster 1(2) as the individual sheets where \jpar\ aligns(anti-aligns) with $\mathbf{B}$, and cluster 3 as the background plasma. 
The robustness of the reported clusters strongly indicates that the required input properties are universal and set by the physics of the turbulence.

\section{Statistical Distributions of SOM clusters} \label{app:distribution}

\begin{figure*}
    \centering
    \includegraphics[width=0.9\linewidth]{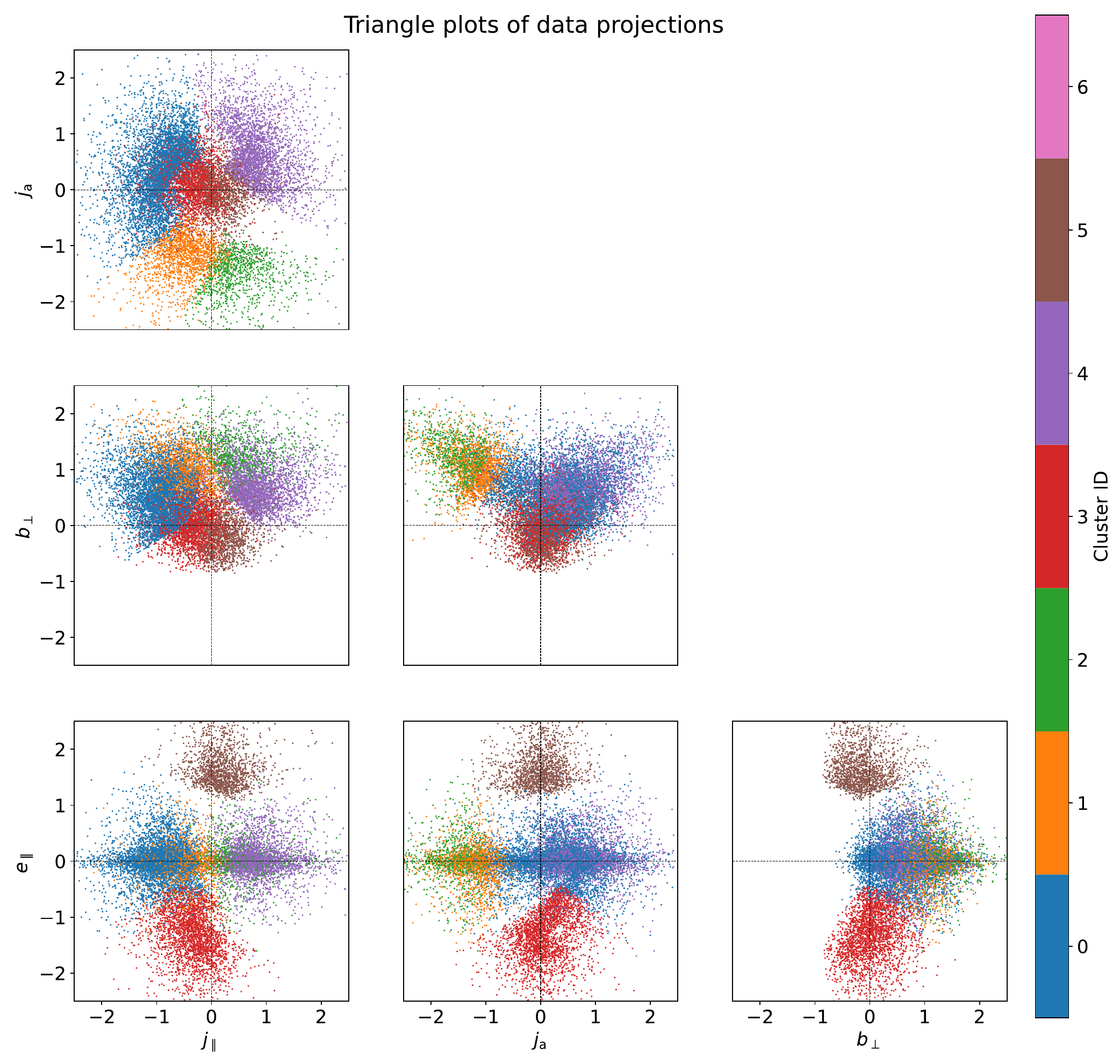}
    \caption{Feature-to-feature scatter plots of SOM clusters from Figure~\ref{fig:cluster_umap}. For clarity, cluster 6 has been excluded, and a random sample of $10^5$ points ($\sim 5\% \,L^3$) is drawn for each plot. Each data point is color-coded by its assigned cluster label.}
    \label{fig:triangle}
\end{figure*}

\begin{figure*}
    \centering
    \includegraphics[width=0.9\linewidth]{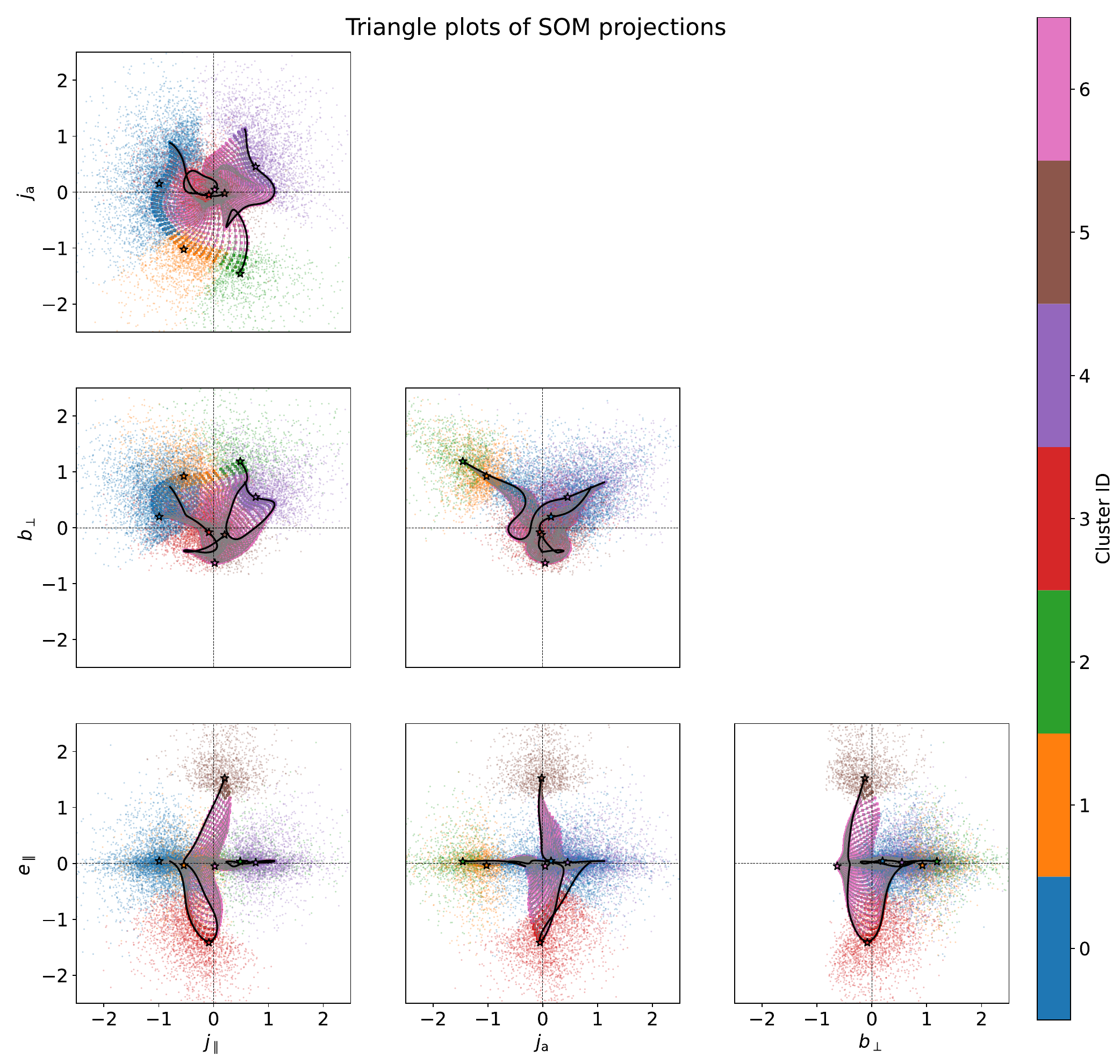}
    \caption{Scatter plots of SOM clusters (in the background; similar to Figure~\ref{fig:triangle}) and projections of the SOM lattice (in the foreground), where each square is a node, and the stars are locations of the centroids of each cluster. 
    The gray lines represent connected nodes along the $X$-axis in Figure~\ref{fig:cluster_umap}. 
    The two black curves show the horizontal boundaries of the map with $Y = 0$\ and $Y = 36$.}
    \label{fig:lattice_triangle}
\end{figure*}

\begin{figure*}
    \centering
    \includegraphics[width=0.7\linewidth]{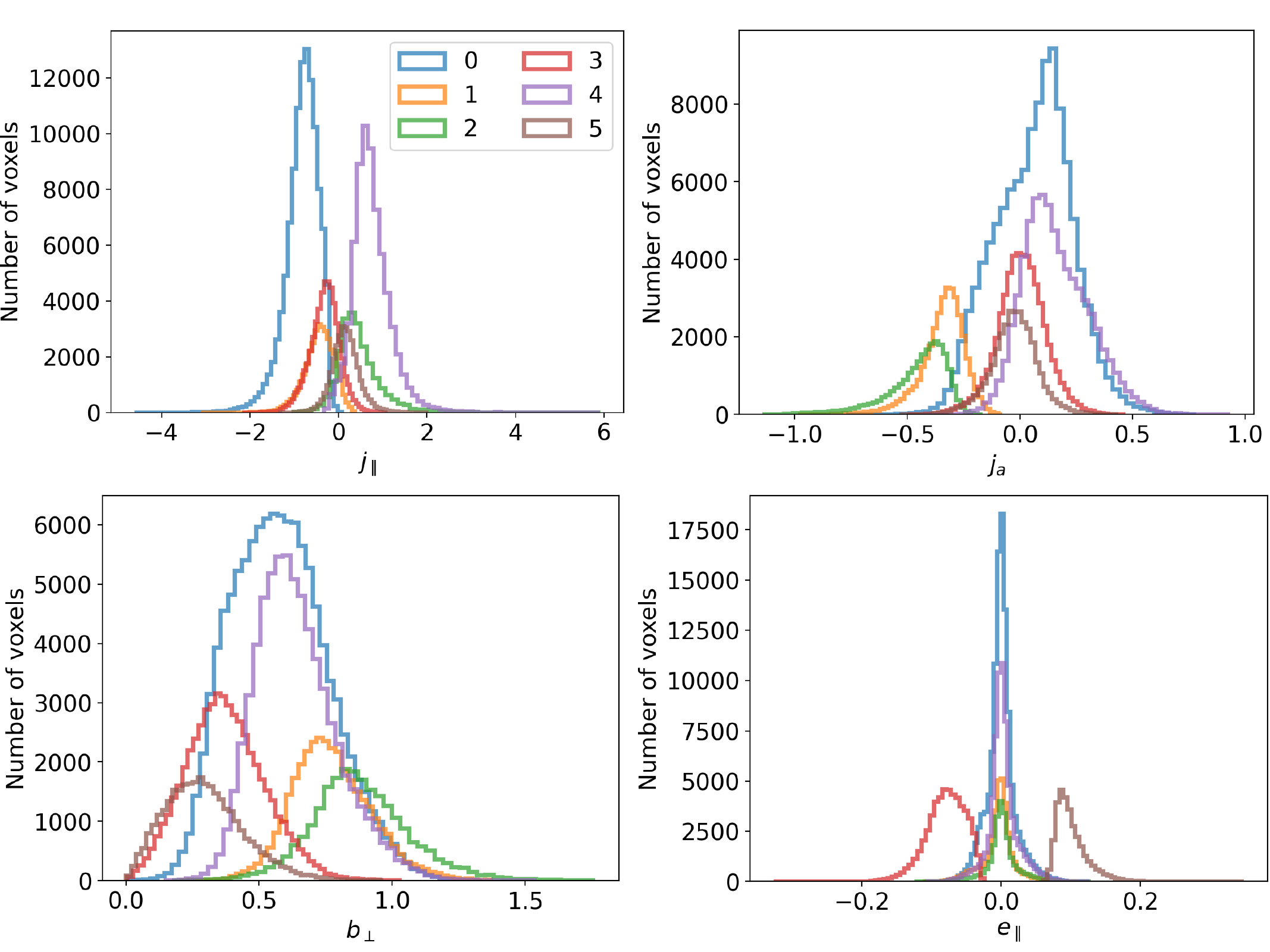}
    \caption{Histograms of each feature used in training, separated by clusters from Figure~\ref{fig:cluster_umap}. For clarity, cluster 6 has been omitted.}
    \label{fig:histograms}
\end{figure*}

We perform several statistical analyses based on the fiducial SOM clusters to further explore the physics behind each cluster. For each of these analyses, we omit cluster 6, which only contains the background plasma. 

Figure~\ref{fig:triangle} shows scatter plots between pairs of features used. A random sample of $10^5$\ data points is used, and cluster 6 (background) is omitted, to improve the clarity of the plots. The data points are quite clearly separated in a multi-dimensional space. For instance, clusters 0 (blue) and 4 (purple), as well as clusters 1 (orange) and 2 (green), are separated along the \jpar\,$=0$\ line. As we established in the main text, these clusters are either parts of the double sheet structures (clusters 1 and 2), or they are individual current sheets (clusters 0 and 4). Another correlation that we can see is in the \epar\ plots, where clusters 3 (red) and 5 (brown) are on opposite tail ends of the distribution. We also observe that the double current sheets tend to have markedly stronger \bperp\ than the rest of the data, pointing to the prominent peak in $B_{p1}$~seen in Figure~\ref{fig:single_double}.
Figure~\ref{fig:lattice_triangle} shows the same scatter plots as Figure~\ref{fig:triangle}, but with the projections of the SOM lattice plotted in the foreground, showing the non-linear topology of the lattice. 
It can be used to read the non-linear combination of the feature-space parameters that the map uses for the clustering:
the major axis (i.e., slices along $X$) is mainly oriented along the $j_a$ feature but also captures variations in $e_\parallel$ in the top part of the map (for $Y \gtrsim 30$).
The minor axis (i.e., slices along $Y$) is somewhat oriented along $j_\parallel$ and $b_\perp$.
However, we emphasize that the real map is composed of a non-linear combination of the features and is, therefore, highly tangled in the feature space.

Figure~\ref{fig:histograms} shows histograms of each feature in our fiducial SOM run, separated by the clusters seen in Figure~\ref{fig:cluster_umap}. These histograms support the same conclusions as the scatter plots from Figure~\ref{fig:triangle}. 

\section{Additional simulation data} \label{app:other-sims}

We compose a set of three 3D fully kinetic PIC simulations to study the formation and morphology of current sheets. The main text explores the details and results of the decaying turbulence box simulation. 
Here, we investigate two other fully kinetic PIC simulations with similar conclusions as in the main text.

\subsection{Reconnecting Harris current sheet}

\begin{figure}[ht]
    \centering
    \includegraphics[height=0.95\textheight]{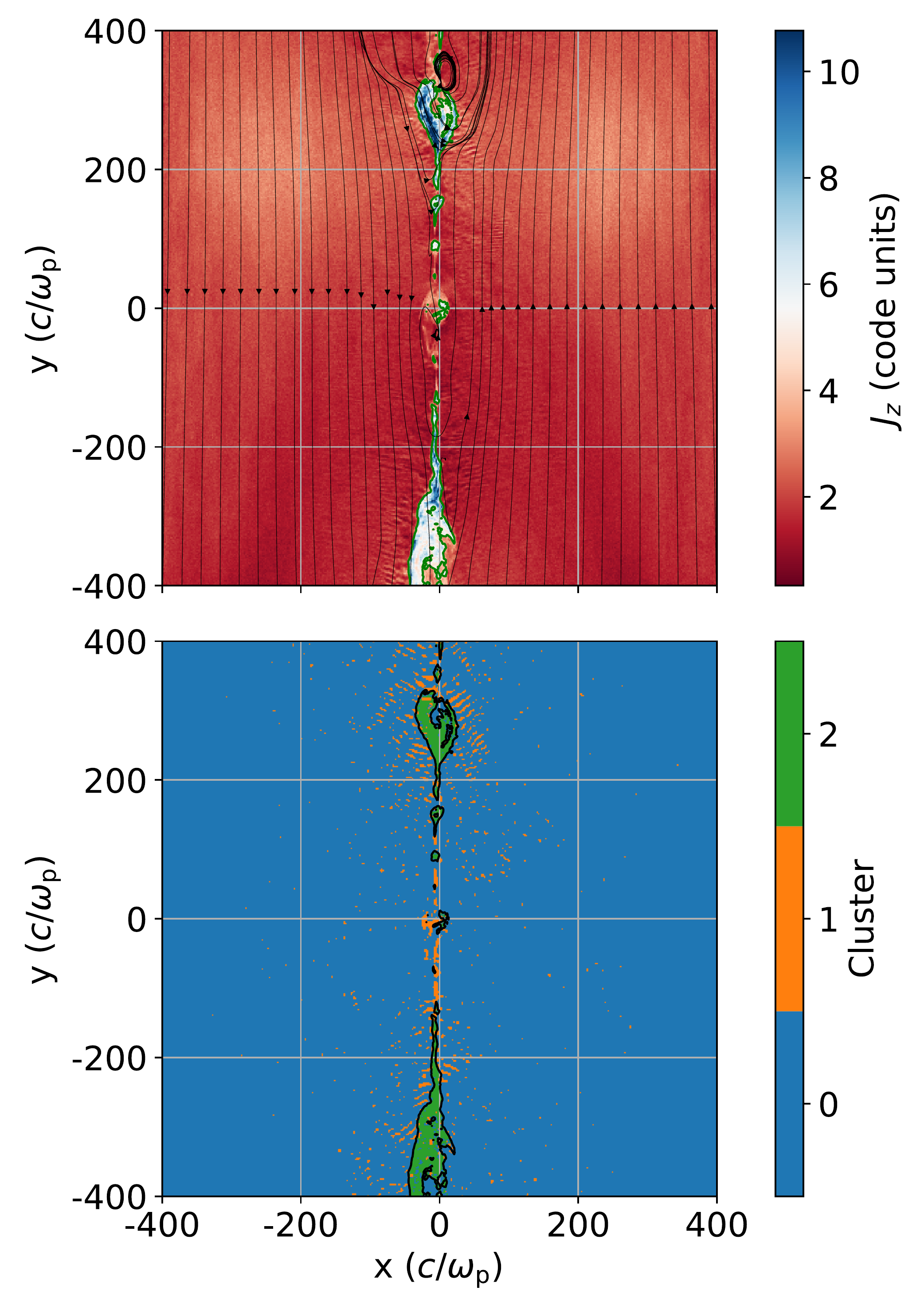}
    \caption{$xy$-slice of current density along z-axis ($J_z$, top) vs. \awesom~clustering result (bottom) at $z = -40 c/\omega_{\rm p}$ for the Harris current sheet simulation. Streamlines in the top panel show in-plane magnetic field lines. Contours in both panels show regions where $J_z > 2 J_{\rm rms}$, representing the boundary of the Harris current sheet.}
    \label{fig:fig_harris}
\end{figure}

We use data from \citet{zhang_origin_2023} for an ideal simulation of a Harris current sheet. 
The magnetic field $B_0$ is initialized to reverse from $+y$~to $-y$~across a current sheet at $x = 0$. The magnetization is $\sigma = 10$; a guide field of $B_g = 0.1 B_0$~along $z$~also presents. They initialize a cold $e^\pm$-pair plasma with rest-frame density, $n_0$, of 2 particles per cell per species. 
Fresh plasma and magnetic flux are continuously injected along the $x$~direction of inflow. 
The simulation domain covers $L_x = L_y = 2 L_z = 1600 c/\omega_{\rm p}$. 
For ease of analysis, we select a cubic region of size $L = L_z = 800 c/\omega_{\rm p}$, centered on $x=y=z=0$.

We apply \awesom\ to the Harris sheet data with these parameters: $N = 10^6$, $X = 70$, $Y = 49$, and $\alpha_0 = 0.1$. 
We use three features to train the model: $j_z$, \bperp, and \epar. 
Given the idealized initial conditions, we do not recover the \jasym\ because only one current sheet is formed. 
All other parameters are identical to the fiducial run in the main text.


Figure~\ref{fig:fig_harris} shows a $xy$-slice of an \awesom~clustering result of the Harris sheet. 
In this idealized simulation, cluster 2 (green) has an almost-one-to-one correspondence with a current density threshold of $J_z > 2 J_{\rm rms}$, commonly used as a proxy for identifying current sheets in simulations \citep{zhdankin_statistical_2013}. 
Cluster 0 (blue) accounts for the rest of the background plasma, and cluster 1 (orange) are locations with strong \epar.

\subsection{Driven turbulence}

\begin{figure}[ht]
    \centering
    \includegraphics[height=0.95\textheight]{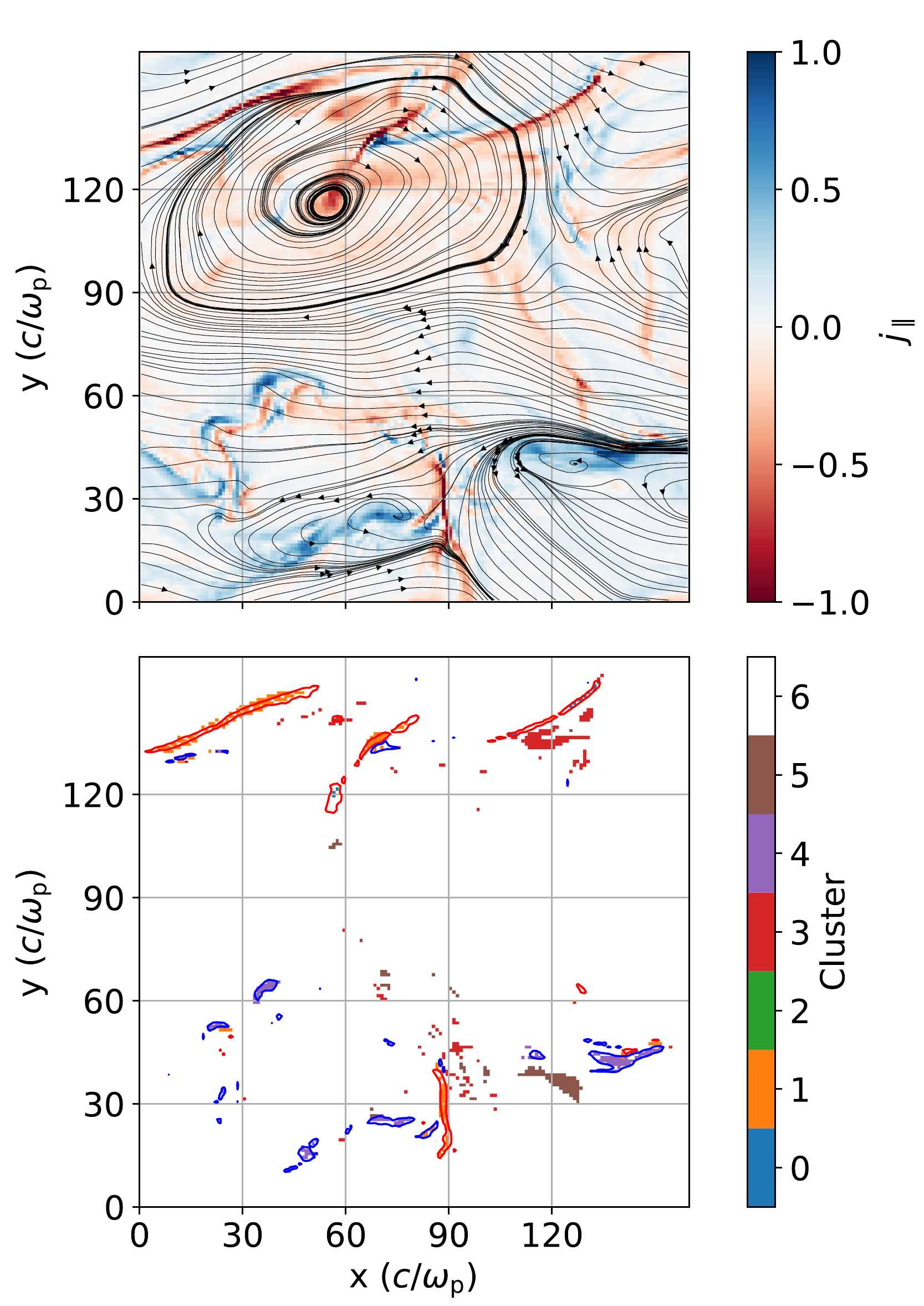}
    \caption{$xy$-slice of \jpar~(top) vs. \awesom~clustering result (bottom) at $z = 120 c/\omega_{\rm p}$. Streamlines in the top panel show the in-plane magnetic field, $B_{xy}$. Red and blue contours in the bottom panel show locations where $j_{\parallel} < -3 \, j_{\parallel, \rm rms}$, and where $j_{\parallel} < -3 \, j_{\parallel, \rm rms}$, respectively. The color of the clusters have the same physical meaning as in Figure~\ref{fig:merged_double}.}
    \label{fig:fig_driven}
\end{figure}

We also use snapshots from a continuously driven turbulence simulation \citep{nattila_radiative_2024}.
Similar to the freely-evolving turbulence case, the domain is a triply periodic cubic box of length $L = 640 c/\omega_{\rm p}$, which contains a neutral $e^\pm$-pair plasma with 8 particles per cell per species. 
The domain has a magnetization of $\sigma = 10$ and is initialized with a uniform magnetic field $\mathbf{B}_0 = B_0 \hat{\mathbf{z}}$. 
The turbulence is continuously excited by driving an external current with an oscillating Langevin antenna formalism \citep{tenbarge_oscillating_2014}. 
The antenna has the following properties: driving scale $l_0 = L/2$, amplitude $\delta B / B_0 = 0.8$, frequency $\omega_{\rm ant}/\omega_{0} = 0.8$ (where $\omega_0 = 2\pi/t_0$ is the eddy-turnover frequency, and $t_0 = l_0/v_{\rm A}$ is the eddy-turnover time), and a decorrelation time of $\omega_{\rm dec} / \omega_0 = 0.6$.
The SOM presented below was trained on one snapshot of the simulation, taken at $t \approx 1.5\;l_0/c$, which exhibits a magnetic power spectrum  $\propto k_\perp^\alpha$ with a slope of $\alpha \approx -2$.

We downsample the dataset by a factor of four in each dimension such that $\Delta x = 4 c/\omega_{\rm p}$ and trained \awesom\ on the resulting data. 
We set $N = 160^3$, $X = 91$, $Y = 37$, and $\alpha_0 = 0.1$ in \awesom.
Four features are used, similar to the freely evolving turbulence case: \jpar, \jasym, \bperp, and \epar.

Figure~\ref{fig:fig_driven} shows a $xy$-slice plot of this simulation with a SOM~realization after separation of single and double current sheets, as detailed in Section~\ref{subsec:separate}.
In general, the intermittent structures in this domain are identified by \awesom. 
Cluster 0 (blue) and cluster 4 (purple) are single current sheets that are aligned and anti-aligned with $\mathbf{B}$, respectively. 
Cluster 1 (orange) are double current sheets. 
Meanwhile, cluster 3 (red) and cluster 5 (brown) are locations where \epar\ is high, and cluster 6 (white) are locations of the background plasma.
The volume-filling fraction of each cluster-of-interest (for clusters $k = $ 0, 1, 2, ...) is: $n_{{\rm fill},\ k} \approx 0.09\%, 0.91\%, 0\%, 0.70\%, 0.37\%, 0.57\%, {\rm and}\ 97.36\%$.

Notably, the volume-filling fraction of the sheets is much smaller than that found in the freely-evolving turbulence simulation, $n_{\rm fill,0+1+4} \approx 1.37\%$. 
Given the similar (down-sampled) resolution of both the decaying turbulence and driven turbulence box, such a discrepancy in volume-filling fractions may be explained by the different modes of perturbation, or by the size of the turbulence driver ($l_0 = L/2$, vs. $L/4$\ in the decaying turbulence simulation). 
Alternatively, the difference seen here could be due to a transient turbulent effect such as a flare, that is inadequately captured by individual snapshots.
The largest current sheets in the continuously driven turbulent domain could be sourced by the external driving mechanism, rather than the non-linear turbulent cascade, and, hence, we do not expect the double sheets to be as prominent in this simulation compared to the fiducial case presented in the main text.
Nevertheless, the analysis shows qualitatively similar results to the freely-evolving turbulence simulations.

\bibliography{main}{}
\bibliographystyle{aasjournal}


\end{document}